\documentclass[twocolumn,showpacs,superscriptaddress]{revtex4}
\usepackage{graphicx}
\usepackage{amsmath}
\usepackage{amsfonts}
\usepackage{amssymb}

\newcommand{\HF}{{\mathrm{HF}}}
\newcommand{\Tr}{{\mathop{\rm{Tr}}\nolimits\,}}
\newcommand{\br}{{\bf r}}

\begin{document}
\title{Proposed definitions of the correlation energy density from a
Hartree-Fock starting point: The two-electron Moshinsky model atom as an exactly
solvable model}
\author{N. H. March}
\affiliation{Department of Physics, University of Antwerp,\\
Groenenborgerlaan 171, B-2020 Antwerp, Belgium}
\affiliation{International Centre for Theoretical Physics,\\
Strada Costiera, 11, Miramare, Trieste, Italy}
\affiliation{Oxford University, Oxford, UK}
\author{A. Cabo}
\affiliation{International Centre for Theoretical Physics,\\
Strada Costiera, 11, Miramare, Trieste, Italy}
\affiliation{Grupo de F\'{\i}sica Te\'orica, Instituto de Cibern\'etica,
Matem\'atica y F\'{\i}sica,\\ Calle E, No. 309, Vedado, La Habana, Cuba}
\author{F. Claro}
\affiliation{International Centre for Theoretical Physics,\\
Strada Costiera, 11, Miramare, Trieste, Italy}
\affiliation{Facultad de F\'{\i}sica, Pontificia Universidad Cat\'{o}lica de
Chile,\\ Campus San Joaquin, Santiago de Chile, Chile}
\author{G. G. N. Angilella}
\email[Corresponding author. E-mail: ]{giuseppe.angilella@ct.infn.it}
\affiliation{Dipartimento di Fisica e Astronomia, Universit\`a di
Catania,\\ and CNISM, UdR Catania, and INFN, Sez. Catania,\\ 64, Via S. Sofia,
I-95123 Catania, Italy}

\begin{abstract}
In both molecular physics and condensed matter theory, deeper understanding of
the correlation energy density $\epsilon_{c}(\mathbf{r})$ remains a high
priority. By adopting L\"owdin's definition of correlation energy as the
difference between the exact and the Hartree-Fock values, here we propose two
alternative routes to define this. One of these involves both exact and
Hartree-Fock (HF) wavefunctions, while the second requires a coupling constant
integration. As an exact analytical example of the first route, we treat the
two-electron model atom of Moshinsky, for which both confinement potential and
interactions are harmonic. Though the correlation energy density
$\epsilon_{c}(\mathbf{r})$ is known analytically, we also investigate
numerically its relation to the exact ground-state density in this example.\\
\pacs{%
31.15.Ew,
31.25.-v,
31.25.Eb
}
\end{abstract}
\maketitle

\section{Introduction}

It is true to say that one of the remaining problems in both molecular physics
and condensed matter theory is to gain deeper understanding of the correlation
energy density. We have recently approached this problem via M{\o}ller-Plesset
(MP) perturbation theory \cite{Cabo:06,Grassi:07}. By adopting L\"owdin's
\cite{Loewdin:55a} definition of the ground-state correlation energy as the
difference between the exact and the Hartree-Fock (HF) values, here we shall
propose two, formally exact, routes to the correlation energy density. The
first of these, as in \cite{Grassi:07}, starts out from the so-called level-shift
formula \cite{March:67}, but in contrast to \cite{Grassi:07}, where low-order
perturbation theory is invoked, our central example is exactly solvable, which
means that the MP series has been summed to all orders. This example is the
model two-electron atom introduced by Moshinsky \cite{Moshinsky:68} and it is
therefore natural enough that we pose the two-electron atom example formally
exactly in section~\ref{sec:els} immediately below. Section~\ref{sec:euls}
presents the exact theory for the Moshinsky model. Comparison is made in
section~\ref{sec:densdep} of the total kinetic energy, including correlation of
the Moshinsky atom with that of the (non-relativistic) He-like series of atomic
ions for large atomic number. An alternative route for defining the correlation
energy density is then proposed in section~\ref{sec:green}, which may prove to
come into its own in solid-state theory rather than molecular physics.
Section~\ref{sec:summary} constitutes a summary, plus proposals for future
studies which should be fruitful.

\section{Exact level-shift theory for the ground state of atoms and molecules}
\label{sec:els}

Let us consider an $N$-body system described by the Hamiltonian
\begin{subequations}
\begin{eqnarray}
\label{hamil}
H  &=& H_{0}+H_{I},\\
H_{0}  &=& \int d\mathbf{r}_{1}\,d\mathbf{r}_{2}\Psi^{\dag}(\mathbf{r}_{1})
h (\mathbf{r}_{1},\mathbf{r}_{2})\Psi(\mathbf{r}),\\
H_{I}  &=& \frac{1}{2}\int d\mathbf{r}_{1}\,d\mathbf{r}_{2}  
\Psi^{\dag}(\mathbf{r}_{1})\Psi^{\dag}(\mathbf{r}_{2}) v(\mathbf{r}_{1}
,\mathbf{r}_{2})\Psi(\mathbf{r}_{2})\Psi(\mathbf{r}_{1})  ,
\end{eqnarray}
\end{subequations}
where the field operators in second quantization are defined as usual by their
expressions in terms of the creation and annihilation operators and their
standard commutation relations:
\begin{subequations}
\begin{eqnarray}
\Psi(\mathbf{r})  &=& \sum_{k}\Psi_{k}(\mathbf{r})\,a_{k},\\
\Psi^{\dag}(\mathbf{r})  &=& \sum_{k}\Psi_{k}^{\ast}(\mathbf{r})\,a_{k}^{\dag},\\
\relax[  a_{k},a_{k^{\prime}}^{\dag}] _{+}  &=& \delta_{kk^{\prime}},
\quad\left[  a_{k},a_{k^{\prime}}\right] _{+}=0.
\end{eqnarray}
\end{subequations}
Here,  $v(\mathbf{r}_{1},\mathbf{r}_{2})$ is the interaction potential between
the particles and $h(\mathbf{r}_{1},\mathbf{r}_{2})$ is the kernel of a
one particle operator that reduces to the kinetic energy operator in the usual
cases. However, it may also embody the effect of an external potential and
other effects. The creation (annihilation) operator $a^{\dag}_k$ ($a_k$)
is assumed to create (annihilate) particles in states described by the
wave-functions $\Psi_{k}(\mathbf{r})$. Here, as is usual, $\mathbf{r} \equiv
(\mathbf{x},s)$, with $\mathbf{x}$ denoting the particle 
position, and $s$ the particle spin, and $k$ is the collective quantum
number associated with the basis states $\left\{  \Psi_{k}\right\}$. We
will assume  that the free Hamiltonian is diagonal in the spin variable  and
also that the interaction is  spin-independent, \emph{i.e.}
\begin{subequations}
\begin{eqnarray}
\label{eq:hh1}
h(\mathbf{r}_{1},\mathbf{r}_{2})  &=& h({\mathbf{x}}_{1},{\mathbf{x}}_{2})
\delta_{s_{1}s_{s}}\\
v(\mathbf{r}_{1},\mathbf{r}_{2})  &=& v({\mathbf{x}}_{1},{\mathbf{x}}_{2}).
\end{eqnarray}
\end{subequations}

Following \cite{Grassi:07}, let us use the level-shift formula \cite{March:67},
taking as the unperturbed problem the Fock Hamiltonian $H_\HF$, with ground
state energy $E_{0}$. Then, for two electrons, we can write explicitly for the
correlation energy density $\epsilon_{c}(\mathbf{r}_{1})$
\begin{equation}
\epsilon_{c}(\mathbf{r}_{1})=\frac{\int\Psi^\ast (\mathbf{r}_{1},
\mathbf{r}_{2})[H-H_\HF]\Phi_\HF (\mathbf{r}_{1},\mathbf{r}_{2})
d\mathbf{r}_{2}}{\int\Psi^\ast (\mathbf{r}_{1},\mathbf{r}_{2})\Phi_\HF
(\mathbf{r}_{1},\mathbf{r}_{2})d\mathbf{r}_{1} d\mathbf{r}_{2}}, 
\label{2.1}%
\end{equation}
where $\Psi$ is the exact ground-state wave function. It should be
noted that the definition above of the correlation energy density,
Eq.~(\ref{2.1}), is not unique. However, this circumstance is not
necessarily  problematic, because any alternative definition should lead  to
the same total integrated correlation energy.  This is a similar situation,  and
moreover also seems to be close connected, with the known lack of  precise 
definitions of the energy-momentum tensor for general physical systems.
Therefore, such a property  should not restrict the value and utility of the
concept, whenever  it becomes possible to construct  a theoretical scheme in
which this energy density  plays a relevant role independently  of its non-unique
definition. From Eq.~(\ref{2.1}) we then have, in an obvious notation
\begin{eqnarray}
\langle\Psi |\Phi_\HF \rangle\epsilon_{c}(\mathbf{r}_{1})&=&
E\int
\Psi^\ast (\mathbf{r}_{1},\mathbf{r}_{2})\Phi_\HF (\mathbf{r}_{1},
\mathbf{r}_{2})d\mathbf{r}_{2} \nonumber\\
&-&
E_{0}\int\Psi^\ast (\mathbf{r}_{1},\mathbf{r}_{2})\Phi_\HF
(\mathbf{r}_{1},\mathbf{r}_{2})d\mathbf{r}_{2}, 
\label{2.2}
\end{eqnarray}
where in reaching Eq.~(\ref{2.2}) from Eq.~(\ref{2.1}), $H$ has been allowed to act
on $\Psi^\ast$, and $H_\HF$ on $\Phi_\HF$. Of course, by integrating
both sides of Eq.~(\ref{2.2}) over all $\mathbf{r}_{1}$, we obtain a trivial
identity for $E-E_{0}$, the latter quantity being simply the total L\"owdin
correlation energy $E_{c} = \int\epsilon_{c}(\mathbf{r}) d\mathbf{r}$.

While Eq.~(\ref{2.2}) is of course, valid for the (nonrelativistic) He-like
series of two-electron ions with atomic number $Z$, we do not presently know
$\Psi^\ast$ and $E$ analytically. Therefore in section~\ref{sec:euls}
immediately below, we turn to illustrate Eq.~(\ref{2.2}) analytically by appeal
to the Moshinsky two-electron atom model \cite{Moshinsky:68}.

\section{Exact use of the level-shift formula for the harmonic Moshinsky
two-electron model}
\label{sec:euls}

The Moshinsky model atom has confining (external) potential 
$\frac{1}{2}(|\mathbf{r}_{1}|^{2}+|\mathbf{r}_{2}|^{2})$ and particle interaction also of harmonic
form $\frac{1}{2} k |\mathbf{r}_{1}-\mathbf{r}_{2}|^{2}$. Using coordinates
$\mathbf{R}=(\mathbf{r}_{1}+\mathbf{r}_{2})/\sqrt{2}$ and
$\mathbf{r}=(\mathbf{r}_{1}-\mathbf{r}_{2})/\sqrt{2}$, the exact ground-state
wavefunction takes the form
\begin{equation}
\Psi(\mathbf{r},\mathbf{R})=\frac{(1+2k)^{\frac{3}{8}}}{\pi^{\frac{3}{2}}}%
\exp\left(-\frac{1}{2}\mathbf{R}^{2}\right)\exp\left(-\frac{1}{2}(1+2k)^{\frac{1}{2}}%
\mathbf{r}^{2}\right).
\label{3.1}%
\end{equation}
Less well known is the fact that the corresponding $\Phi_\HF$ wavefunction
entering the key expression (\ref{2.2}) for the correlation energy density has
the exact form \cite{Ballentine:98}
\begin{equation}
\Phi_\HF (\mathbf{r}_{1},\mathbf{r}_{2})=\phi(\mathbf{r}_{1})\phi
(\mathbf{r}_{2}), 
\label{3.2}
\end{equation}
where
\begin{equation}
\phi(\mathbf{r})=\frac{(1+k)^{\frac{3}{8}}}{\pi^{\frac{3}{4}}}\exp\left(-\frac
{1}{2}(1+k)^{\frac{1}{2}}\mathbf{r}^{2}\right). 
\label{3.3}
\end{equation}
It is understood that Eq.~(\ref{3.2}) only provides the spatial dependence of
the ground state, whose overall antisymmetric character is to be provided by the
spin dependence. Since the spin structure of the ground state problem is fixed
by its singlet character, we can safely employ the symbol $\mathbf{r}$ below to
denote only the spatial coordinates.

Plots of the ``overlap'' $\langle\Psi|\Phi_\HF\rangle$ are already available as
functions of the particle-particle interaction strength $k$: \emph{e.g.} for
$k=1$ the overlap is 0.94. Inserting Eqs.~(\ref{3.1}) and (\ref{3.2}) into
the right-hand-side of Eq.~(\ref{2.2}) we obtain an exact result for
$\epsilon_{c}(\mathbf{r})$, now spherically symmetric in the Moshinsky atom
model. The correlation energy density, the HF, and exact electron densities,
respectively, have all the analytic form
\begin{subequations}
\begin{align}
\epsilon_{c}(\mathbf{r})  &= E_{c}\left(\frac{\alpha}{\pi}\right)^{\frac{3}{2}}%
\exp(-\alpha r^{2}), & \alpha &=\frac{b^{2}-(1-a)^{2}}{4b} ,
\label{3.4}\\
\rho(\mathbf{r})  &= 2\left(\frac{\beta}{\pi}\right)^{\frac{3}{2}}\exp(-\beta r^{2}),
& \beta &= \frac{2a}{1+a},
\label{3.5}\\
\rho_\HF (\mathbf{r})  &=2\left(\frac{\gamma}{\pi}\right)^{\frac{3}{2}}\exp
(-\gamma r^{2}),  & \gamma&=\sqrt{1+k} ,
\label{3.6}\\
a  &=\sqrt{1+2k}, & b&= 1+a+2\gamma.
\label{3.7}%
\end{align}
\end{subequations}

\begin{figure}[t]
\centering
\includegraphics[width=0.9\columnwidth]{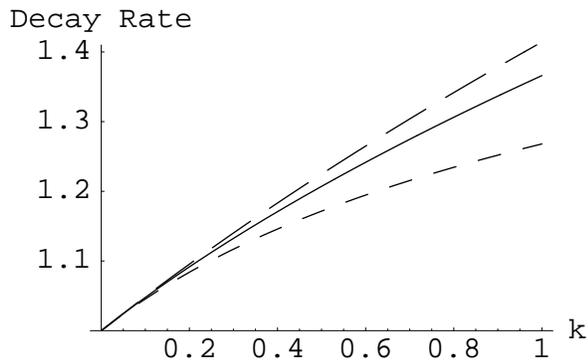}
\caption{Decay rate for the correlation energy density ($\alpha$, full line),
and the exact ($\beta$, short dashed) and Hartree-Fock ($\gamma$, long dashed)
one-particle densities, in terms of the coupling parameter $k$ (All
quantities are in atomic units).}
\label{grafico1}
\end{figure}

\begin{figure}[t]
\centering
\includegraphics[width=0.9\columnwidth]{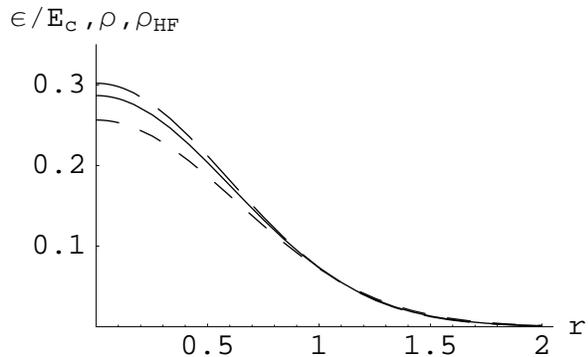}
\caption{Radial dependence of the reduced correlation energy density (full
line), exact (short dashed) and Hartree-Fock (long dashed) particle
densities. Here, $k=1$. All quantities are in atomic units.}
\label{grafico2}
\end{figure}

Notice that all three functions are gaussians, albeit with a different decay
rate exponent. Figure~\ref{grafico1} exhibits the $k$-dependence of the latter.
At $k=0$ the interaction vanishes, and all three coefficients equal one. As the
interaction is turned on they increase, departing slowly from each other.
Figure~\ref{grafico2} shows the reduced correlation energy density
$\epsilon_{c}(\mathbf{r})/E_{c}$ (full line) together with the HF (long dashes)
and exact (short dashes) densities normalized to one, for $k=1$. It is apparent
from the functional identity and the weak parameter divergence that there is an
intimate connection between the correlation energy density and the exact or HF
electron densities. This result is compatible with the definition of
$\epsilon_{c}(\mathbf{r})$ as proportional to the HF one particle density, as
proposed in \cite{Grassi:07}. Exact relations for the Moshinsky model are presented
in the Appendix.

\section{Density dependence of the total kinetic energies in the Moshinsky
model atom and the H\lowercase{e}-like atomic ions at large atomic number $Z$}
\label{sec:densdep}

Let us illustrate in this section two examples of physical systems in which the
total kinetic energy can be expressed as a functional of the density. We note
first that the total kinetic energy of the Moshinsky model treated in
section~\ref{sec:euls} above can be expressed exactly in terms solely of $\rho
(\mathbf{r})$. From \cite{Holas:03} one knows that the kinetic energy density $t$
now defined from the wavefunction form $(\nabla\Psi)^{2}$ is given by
\begin{equation}
t_{\mathrm{Mosh}}(\mathbf{r})=\frac{1}{2}\rho(\mathbf{r})\left[  \frac{3}{2}
\frac{(d-1)^{2}}{d}-\frac{2d-1}{d} \log \frac{\rho(\mathbf{r})}{\rho
(\mathbf{0})}\right]  ,
\end{equation}
where
\begin{equation}
d^{-1}=2-\pi\left[\frac{\rho(\mathbf{0})}{2}\right]^{\frac{2}{3}} .
\label{alfa}
\end{equation}
As noted earlier, in the Moshinsky atom the ground state $\rho(\mathbf{r})$ is
purely Gaussian. Notice that, in contrast to the He-ions at large $Z$, to be
discussed below, there is a $\log\rho(\mathbf{r})$ term. Because everything is
harmonic we expect that the potential energy $U$ will coincide with the kinetic
term $T$ and that both will be a half of the total energy $E$, that is
\begin{equation}
T=U=\frac{E}{2}.
\end{equation}
Thus the correlation energy density can be obtained as
\begin{equation}
E_{\mathrm{corr}}\simeq 2T_{\mathrm{corr}}=2\int
t_{\mathrm{corr}}(\mathbf{r})d\mathbf{r}.
\end{equation}
Here, $t_{\mathrm{corr}} (\mathbf{r}) = t_{\mathrm{Mosh}} (\mathbf{r}) -
\epsilon_{\mathrm{HF}} (\mathbf{r})/2$, and $\epsilon_{\mathrm{HF}}
(\mathbf{r})/2$ is defined as a  density of HF energy constructed from one of 
its  expressions as spatial integrals. In such a way, the correlation energy
density is defined as $\epsilon_c (\mathbf{r}) =2 t_{\mathrm{corr}}
(\mathbf{r})$.

\subsubsection{Kinetic energy, including correlation, of the He-like atomic
ions with large atomic number}

Following the work of Schwartz \cite{Schwartz:59} on the He-like atomic
ions with large $Z$ (however, still non-relativistic)  the total kinetic
energy $T$ has been obtained by Gal, March and Nagy \cite{Gal:99} as
\begin{equation}
T=-\frac{1}{2}\left[  \frac{\rho^{\prime}(\mathbf{r})}{\rho(\mathbf{r})}
\right]_{\mathbf{r=0}}\int\frac{\rho(\mathbf{r})}{|\mathbf{r|}}
d\mathbf{r-}\frac{1}{8}\left[  \frac{\rho^{\prime}(\mathbf{r})}
{\rho(\mathbf{r})}\right]_{\mathbf{r}=0}^{2}\int\rho(\mathbf{r})d\mathbf{r}.
\end{equation}
This expression for the total kinetic energy, again including kinetic
correlation energy is evidently determined therefore solely by $\rho
(\mathbf{r})$ and the derivative of $\log\rho(\mathbf{r})$. Thus, the
total kinetic energy density $t(\mathbf{r})$ takes the form
\begin{equation}
t(\mathbf{r})=-\frac{1}{2}\left[  \frac{\rho^{\prime}(\mathbf{r})}
{\rho(\mathbf{r})}\right]_{\mathbf{r=0}}\frac{\rho(\mathbf{r})}
{|\mathbf{r|}}\mathbf{-}\frac{1}{8}\left[  \frac{\rho^{\prime}(\mathbf{r}
)}{\rho(\mathbf{r})}\right]_{\mathbf{r}=0}^{2}\rho(\mathbf{r}) .
\end{equation}
In closing this section we emphasize that in contrast to the Moshinsky atom,
the total kinetic energy including correlation now depends on both
$\rho(\mathbf{r})$ and $\nabla\rho(\mathbf{r})$.

\section{Differential second-order density matrix formula for correlation energy
density via a coupling constant integration}
\label{sec:green}

The differential level shift formula (\ref{2.1}) for the correlation energy
density $\epsilon_{c}(\mathbf{r})$ is entirely appropriate for two-electron
systems like the Moshinsky atom. But repeated volume integrations make it
unwieldy for $N$-electron problems, with $N>2$. Therefore, in this section we
derive an alternative route via a coupling constant integration. This leads to a
formula for the L\"owdin correlation energy density $\epsilon_{c}(\mathbf{r})$
characterized by second-order density matrices. Since the
development of efficient methods for the evaluation of second-order density
matrices is in rapid progess, the reduced  number of integrals to be evaluated
can represent a helpful technical advancement, once the expressions for the
density  matrices are already at hand.

\subsection{Introductory example of uniform electron liquid}

To point the way, let us consider the homogeneous electron liquid (HEL). As
emphasized in early work, one of us \cite{March:58,March:59a} has used as the
`coupling constant' the mean interelectronic separation $r_s$. A variant of
Hellmann-Feynman theorem enables to express the ground-state energy for the HEL
(see \emph{e.g.} \cite{Giuliani:05} for a review). In particular, the
ground-state energy per electron, $E(r_s )$, satisfies the virial theorem
\cite{March:58}
\begin{equation}
2T(r_s ) + U(r_s ) = -r_s \frac{dE}{dr_s}
\label{eq:virial}
\end{equation}
where $T$ and $U$ represent kinetic and potential contributions, respectively. As
shown by March and Young \cite{March:59a}, removing $T=E-U$ from
Eq.~(\ref{eq:virial}) allows the resulting first-order differential equation for
$E(r_s )$ to be integrated to yield
\begin{equation}
E(\lambda) = -\lambda \int^\lambda \frac{U(\lambda)}{\lambda^2} d\lambda ,
\end{equation}
where $\lambda = r_s^2$ plays the role of a coupling constant adiabatically
connecting the unperturbed and the exact Hamiltonians, as in Hellmann-Feynman
theorem. But it is well known that $E_\HF (r_s ) = (A/r_s^2 ) - (B/r_s )$, with
$A=\frac{3}{5}$ and $B=\frac{3}{2\pi} \left( \frac{9\pi}{4} \right)^{1/3}$.
Hence $E(\lambda) - E_\HF$ is known via a coupling constant
integration. In HEL, $U$ is determined by the diagonal element of the
second-order density matrix, and hence we seek next a generalization of such a
formula for the L\"owdin correlation energy for an $N$-electron system without
the translational invariance of the HEL.

\subsection{Coupling constant formula}

Motivated by the above HEL example, we have gone back to the treatment by
Stanton \cite{Stanton:62} of the L\"owdin correlation energy $E-E_\HF$ in terms
of such a coupling constant integration. We write the Hamiltonian as
\begin{equation}
H(\lambda) = H_\HF + \lambda H_1 ,
\end{equation}
where $0\leq \lambda \leq 1$ is a coupling constant adiabatically connecting the
HF Hamiltonian to the exact Hamiltonian $H\equiv H(\lambda=1)$. Stanton
\cite{Stanton:62} then generalizes Hellmann-Feynman's theorem to embrace the
case of an $N$-particle Hamiltonian. Beginning with the elementary identity
$E-E_\HF \equiv E(1) -E(0) = \int_0^1 (dE(\lambda)/d\lambda) d\lambda$, Eq.~(12)
in Ref.~\cite{Stanton:62} reads, in obvious notation,
\begin{equation}
E-E_\HF = \int_0^1 
[\langle \Psi(\lambda) | \frac{\partial H}{\partial\lambda} | \Psi(\lambda)\rangle -
\langle \Phi_\HF | \frac{\partial H}{\partial\lambda} |
 \Phi_\HF  \rangle ]
d\lambda ,
\label{eq:Stanton}
\end{equation}
where $\Psi(\lambda)$ is the ground-state eigenvector of the exact Hamiltonian
$H(\lambda)$ with eigenvalue $E(\lambda)$, and $\Phi_\HF$ is the HF ground-state
with energy $E_\HF$. The scalar products $\langle\ldots\rangle$ obviously imply
an integration over the coordinates of $N$ particles. The important point to
stress here is that instead of the quantum-mechanical average
$\langle\Psi|H|\Phi_\HF \rangle$ entering the level shift formula,
Eq.~(\ref{2.1}), Eq.~(\ref{eq:Stanton}) involves `symmetric' averages like
$\langle \Psi | H | \Psi\rangle$ and $\langle \Phi_\HF  | H | \Phi_\HF \rangle$.
These, of course, are achieved at the cost of the coupling constant integration.
But since $H(\lambda=1)$ involves only one- and two-body operators, all the
volume integrations but two for an $N$-electron atom, molecule of cluster can be
achieved by use of the second-order density matrix $\Gamma(\br_1 ,\br_1^\prime ;
\br_2 , \br_2^\prime )$ defined by \cite{Loewdin:55a}
\begin{widetext}
\begin{equation}
\Gamma(\br_1 ,\br_1^\prime ; \br_2 , \br_2^\prime ) =
\frac{N(N-1)}{2} \int \Psi^\ast (\br_1 , \br_2 ,\br_3 ,\ldots \br_N )
\Psi(\br_1^\prime , \br_2^\prime , \br_3 , \ldots \br_N ) d\br_3 \ldots d\br_N .
\label{eq:Gamma}
\end{equation}
Inserting Eq.~(\ref{eq:Gamma}) into Eq.~(\ref{eq:Stanton}) we hence find
\begin{equation}
E-E_\HF = \frac{2}{N(N-1)} \int_0^1 d\lambda \int d\br_1 d\br_2
\left[ h(\br^\prime_1 , \br^\prime_2 ) - h_\HF (\br^\prime_1 , \br^\prime_2 )
\right] \left.\Gamma(\br_1 ,\br_1^\prime ; \br_2 , \br_2^\prime
;\lambda)\right|_{\br_1^\prime = \br_1 , \br_2^\prime = \br_2} ,
\label{eq:hh}
\end{equation}
where use has been made of the following results:
\begin{subequations}
\begin{eqnarray}
\frac{\partial H(\lambda)}{\partial\lambda} &=& H_1 = H - H_\HF \\
\langle \Phi_\HF | \frac{\partial H(\lambda)}{\partial\lambda} | \Phi_\HF
\rangle &=& \langle \Phi_\HF | H-H_\HF | \Phi_\HF \rangle = 0.
\end{eqnarray}
\end{subequations}
In Eq.~(\ref{eq:hh}), the kernel $h(\br^\prime_1 , \br^\prime_2 )$
associated with the `free' Hamiltonian has been defined in Eq.~(\ref{eq:hh1}),
whereas the expression of the Hartree-Fock kernel $h_\HF (\br^\prime_1 ,
\br^\prime_2$ will be given by Eq.~(\ref{eq:hh2}) below.

The final step, as with the level shift formula Eq.~(\ref{2.1}), is to drop the
volume integration over $\br_2$ and hence to achieve the desired result for the
L\"owdin correlation energy density $\epsilon_c (\br_1 )$ as the coupling
constant integration over second-order density matrices as
\begin{equation}
\epsilon_c (\br_1 ) = \frac{2}{N(N-1)} \int_0^1 d\lambda \int d\br_2
\left[ h(\br^\prime_1 , \br^\prime_2 ) - h_\HF (\br^\prime_1 , \br^\prime_2 )
\right] \left.\Gamma(\br_1 ,\br_1^\prime ; \br_2 , \br_2^\prime
;\lambda)\right|_{\br_1^\prime = \br_1 , \br_2^\prime = \br_2} ,
\label{4.6}
\end{equation}
\end{widetext}
The explicit form of the HF kernel $h_{\mathrm{HF}}
(\mathbf{r}_{1},\mathbf{r}_{2})$
in Eq.~(\ref{4.6}) can be written as
\begin{eqnarray}
h_{\mathrm{HF}}(\mathbf{r}_{1},\mathbf{r}_{2})&=&
h(\mathbf{r}_{1},\mathbf{r}_{2}) \nonumber\\
&&+\delta(\mathbf{r}_{1}-\mathbf{r}_{2})
\int d\mathbf{r}_{3} v(\mathbf{r}_{1},\mathbf{r}_{3})\,
\sum_{k} \Psi_{k}^{\ast}(\mathbf{r}_{3})\Psi_{k}
(\mathbf{r}_{3})  
\nonumber\\
&&-
v(\mathbf{r}_{1},\mathbf{r}_{2})\sum_{k}  \Psi_{k}
(\mathbf{r}_{1})\Psi_{k}^{\ast}(\mathbf{r}_{2})  ,
\end{eqnarray}
where the sum over $k$ runs over the filled orbitals of the mean field
problem. After explicitly writing the spin and spatial dependence, this
expression takes the form
\begin{widetext}
\begin{eqnarray}
h_{\mathrm{HF}}(\mathbf{x}_{1},s_{1};\mathbf{x}_{2},s_{1})&=&
h(\mathbf{x}_{1},\mathbf{x}_{2})\delta_{s_{1}s_{2}}
\nonumber\\
&&+ 
\delta_{s_{1}s_{2}}
\delta(\mathbf{x}_{1}-\mathbf{x}_{2})
\sum_{s_{3}=\pm1}\int d
\mathbf{x}_{3} v(\mathbf{x}_{1},\mathbf{x}_{3})\,\sum_{k}
\Psi_{k}^{\ast}(\mathbf{x}_{3},s_{3})
\Psi_{k}(\mathbf{x}_{3},s_{3}) \nonumber\\
&& -v(\mathbf{x}_{1},\mathbf{x}_{2})
\sum_{k}  \Psi_{k}(\mathbf{x}_{1},s_{1})
\Psi_{k}^{\ast}(\mathbf{x}_{2} ,s_{2}) ,
\label{eq:hh2}
\end{eqnarray}
where the single particle orbitals $\Psi_{k}$ satisfy the HF equations
\begin{equation}
\int d\mathbf{r}_{2}h_{\mathrm{HF}}
(\mathbf{r}_{1},\mathbf{r}_{2})\Psi_{k}(\mathbf{r}_2 )=
\epsilon_{k}\Psi_{k}(\mathbf{r}_1 ).
\end{equation}
Now, the second-quantized  HF Hamiltonian can be expressed as
\begin{subequations}
\begin{eqnarray}
H_{\mathrm{HF}}  &=&
\int d\mathbf{r}_{1}\,d\mathbf{r}_{2}\Psi^{\dag}(\mathbf{r}_{1})
h_{\mathrm{HF}}
(\mathbf{r}_{1},\mathbf{r}_{2})\Psi(\mathbf{r}_{2})-\sum_{k}
\frac{1}{2} v_{k} \nonumber\\
&=& \sum_{k} \epsilon_{k} a_{k^{^{\prime}}}^{\dag} a_{k}-\sum
_{k}\frac{1}{2} v_{k}\\
\epsilon_{k}  &=& h_{k}+ v_{k}\\
h_{k}  &=& \int d\mathbf{r}_{1}\,d\mathbf{r}_{2}
\Psi_{k}^{\ast}(\mathbf{r}_{1})
h(\mathbf{r}_{1},\mathbf{r}_{2})\Psi_{k}(\mathbf{r}_{2}),\\
v_{k}  &=& \sum_{k^\prime}\int d\mathbf{r}_{1}\,d\mathbf{r}_{2}
\Psi_{k}^{\ast}(\mathbf{r}_{1})\Psi_{k^\prime}^{\ast}(\mathbf{r}_{2})
v(\mathbf{r}_{1},\mathbf{r}_{2})
\left(\Psi_{k}(\mathbf{r}_{1})\Psi_{k^\prime} (\mathbf{r}_{2})
-\Psi_{k^\prime}(\mathbf{r}_{1})\Psi_{k}(\mathbf{r}_{2}) \right).
\end{eqnarray}
\end{subequations}
\end{widetext}
It should be noted that the field operator $\Psi(\mathbf{r})$ in the above
expressions are now constructed in terms of the single particle mean field
orbitals.
It should also be recalled that in general the kernel $h_\HF (\br_1^\prime
,\br_2^\prime)$ is not simply given by the Fock
operator. It should also incorporate an additive constant which implements the
property
\begin{equation}
\langle\Phi_\HF| H_\HF |\Phi_\HF \rangle = \langle\Phi_\HF |H|\Phi_\HF \rangle
=E_\HF .
\end{equation}
A possible optional form for $H_\HF$ could be the one introduced in
Ref.~\cite{Cabo:06} in order to propose an improvement of the M\o{}ller-Plesset
perturbative expansion. In this approach the above mentioned additive
constant is not required and the second quantized version of $H_\HF$
becomes  a pure bilinear form in the creation and anihilation operators.

We can, so far, only see a way to evaluate Eq.~(\ref{4.6}) wholly analytically
for the two-electron Moshinsky atom. But we have not pressed the details of that
ourselves, since for this simple two-electron model the level shift formula has
overwhelming advantages over the coupling constant integration formula. But with
recent progress in evaluating correlated two-body density matrices for systems
with $N>2$, we expect Eq.~(\ref{4.6}) to rapidly become the advantageous route
to employ \cite{Mazziotti:06}. Therefore, in App.~\ref{app:green}, we present an
alternative method based on the Green function.

\section{Summary and future directions}
\label{sec:summary}

The two proposals made here for the correlation energy density
$\epsilon_{c}(\mathbf{r}_{1})$\ are embodied in Eqs.~(\ref{2.2}) and
(\ref{4.6}). The first one involves knowledge of the exact many-electron
ground-state function $\Psi$, which of course is generally not available. For
the Moshinsky two-electron atom model, however, both $\Psi$ and its HF
counterpart are known. Our results show that the functional form of the
correlation energy density, HF and exact density is the same, in support of a
simple proportionality expression.

The second proposal has a more general character. An integration over a coupling
constant $\lambda$ involving the second order density matrix must be
accomplished. This alternative seems promising for solid state applications.
However, we delay its analysis for further extensions of this work. The
homogenous electron liquid case of the so-called Sawada Hamiltonian (see
\cite{Sawada:57}) would appear then to afford a promising starting point.

Finally, returning to the level shift, the early analytical work of Schwartz
\cite{Schwartz:59} on He-like atomic ions with large $Z$ referred to in section
4, may provide further insight into the use of the differential level shift
formula (\ref{2.1}). This formula, of course, is readily generalized beyond this
two-electron example, but applications then are likely to involve considerable
computational effort.

\begin{acknowledgments}
The authors gratefully acknowledge support from the ICTP through the Condensed
Matter Section. AC received partial support from the Network on \emph{Quantum
Mechanics Particles and Fields} (Net-35) of the OEA at the ICTP. FC was funded
in part by Fondecyt, Grant 1060650.
\end{acknowledgments}

\appendix

\begin{figure}[t]
\centering
\includegraphics[width=0.7\columnwidth]{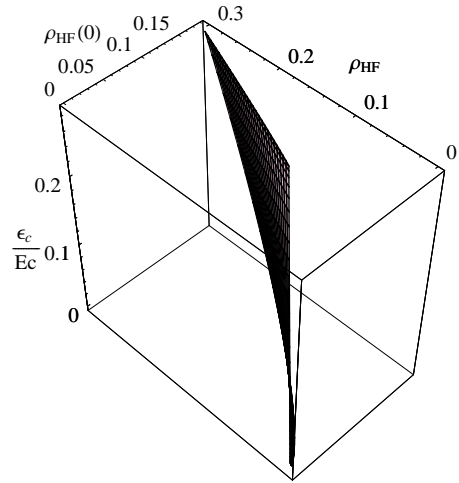}
\caption{Reduced correlation energy density for the Moshinsky model as a
function of the Hartree-Fock density, for a range of values of the latter
evaluated at the origin. All quantities are in atomic units.}
\label{grafico3}
\end{figure}

\section{Dependence of reduced correlation energy on the electron density}

This Appendix is motivated by the simple observation that, for the Moshinsky
model, where we know both the exact electron density $\rho(r)$ and the HF
counterpart $\rho_\HF (r)$, we expect the difference $\rho(r)-\rho_\HF (r)$ to
depend on the reduced correlation energy density
$\epsilon_{c}(\mathbf{r})/E_{c}$.

We have from Eqs.~(\ref{3.4}) and (\ref{3.6}) in the main text, that
\begin{eqnarray}
\rho(\mathbf{r}) &=& \rho(0)\exp(-\beta r^{2}), \label{A1}\\
\rho_\HF (\mathbf{r}) &=& \rho_\HF (0)\exp(-\gamma r^{2}).
\label{A2}
\end{eqnarray}
Hence,
\begin{equation}
\rho(\mathbf{r})-\rho_\HF (\mathbf{r})=\rho(0)(\exp(- r^{2}))^{\beta}
-\rho_\HF (0)(\exp(-r^{2}))^{\gamma}.
\label{A3}
\end{equation}
But we also have
\begin{equation}
\frac{\epsilon_{c}(\mathbf{r})}{E_{c}}=\frac{\epsilon_{c}(0)}{E_{c}}
(\exp(-r^{2}))^{\alpha}.
\label{A4}%
\end{equation}
Hence from Eq.~(\ref{A4}) we can write
\begin{equation}
\exp(- r^{2})=\left(\frac{\epsilon_{c}(\mathbf{r})}{\epsilon_{c}
(\mathbf{0})}\right)^{\frac{1}{\alpha}}.
\label{A5}
\end{equation}
Substituting this equation into Eq.~(\ref{A3}) we find,
\begin{equation}
\rho(\mathbf{r})-\rho_\HF (\mathbf{r})=\rho(0)
\left(\frac{\epsilon_{c}(\mathbf{r})}{\epsilon_{c}
(\mathbf{0})}\right)^{\frac{\beta}{\alpha}}-\rho_\HF (0)
\left(\frac{\epsilon_{c}(\mathbf{r})}{\epsilon_{c}
(\mathbf{0})}\right)^{\frac{\gamma}{\alpha}}.
\label{A6}
\end{equation}
This appears to be a functional relation between 
$\epsilon_{c}(\mathbf{r}) / \epsilon_{c}(\mathbf{0})$ 
and $\rho(\mathbf{r})-\rho_\HF (\mathbf{r})$. 
But of course, on the RHS, the four quantities $\rho(0)$,
$\rho_\HF (0)$,
$\gamma/\alpha$ and $\beta/\alpha$ all
depend on the interaction strength $k$. However, one of these four quantities
is sufficient to carry the fingerprints of $k$ for the others: we single out
therefore $\rho_\HF (0)$. But $\rho(0)$ and $\rho_\HF (0)$ are known from
normalization of the densities to two. Hence we can write,
\begin{equation}
\frac{\rho(0)}{\rho_\HF (0)}=\left(\frac{\beta}{\gamma}\right)^{\frac{3}{2}}.
\label{A7}
\end{equation}
and also
\begin{equation}
\rho(\mathbf{r})=\rho(0)(\exp(-r^{2}))^{\beta}
\label{A8}
\end{equation}
plus
\begin{equation}
\rho_\HF (\mathbf{r})=\rho_\HF (0)(\exp(- r^{2}))^{\gamma}
\label{A9}
\end{equation}
From the latter equations it follows that
\begin{equation}
\rho(\mathbf{r})=\rho_\HF (\mathbf{0})\left(\frac{\beta}{\gamma}\right)^{\frac{3}{2}}
\left(\frac{\rho_\HF(\mathbf{r})}{\rho_\HF(\mathbf{0})}
\right)^{\frac{\beta}{\gamma}}
\end{equation}
Substituting this equation in Eq.~(\ref{A6}) yields
\begin{eqnarray}
\left(\frac{\beta}{\gamma}\right)^{\frac{3}{2}}
\left(\frac{\rho_\HF (\mathbf{r})}{\rho_\HF
(\mathbf{0})}\right)^{\frac{\beta}{\gamma}}-\frac{\rho_\HF (\mathbf{r)}}{\rho_\HF
(\mathbf{0})} &=&\nonumber \\
&& \hspace{-3.5truecm}
\left(\frac{\beta}{\gamma}\right)^{\frac{3}{2}} 
\left(\frac{\epsilon_{c}(\mathbf{r})}{\epsilon_{c}(\mathbf{0})}
\right)^{\frac{\beta}{\alpha}}
-\left(\frac{\epsilon_{c}(\mathbf{r})}{\epsilon_{c}(\mathbf{0})}
\right)^{\frac{\gamma}{\alpha}}.
\label{A11}
\end{eqnarray}
Notice that the strength of the interaction does not appear explicitly in this
formula. Figure~\ref{grafico3} shows a three dimensional plot of
$\epsilon_{c}(\mathbf{r}) / \epsilon_{c}(\mathbf{0})$, $\rho_\HF (\mathbf{r)}$
and $\rho_\HF (\mathbf{0})$, where the latter plays the role of the interaction
strength $k$.

\section{Coupling constant integration in a Green function formula}
\label{app:green}

Following the discussion of Fetter and Walecka \cite{Fetter:04}, we here
attempt to solve the time-independent Schr\"odinger equation for an arbitrary
value of $\lambda$, namely
\begin{equation}
H(\lambda)\Psi(\lambda)=E(\lambda)\Psi(\lambda),
\label{4.2}
\end{equation}
where the wave function $\Psi$ is assumed to be normalized: that is
$\langle\Psi(\lambda)|\Psi(\lambda)\rangle=1$. One immediately finds then from
Eq.~(\ref{4.2}) that
\begin{equation}
E(\lambda)=\langle\Psi(\lambda)|H(\lambda)|\Psi(\lambda)\rangle.
\label{4.3}
\end{equation}
Differentiating with respect to the parameter $\lambda$ yields
\begin{equation}
\frac{d}{d\lambda}E(\lambda)=\langle\Psi(\lambda)|
H_{1}|\Psi(\lambda)\rangle,
\end{equation}
where $H_{1}=H(\lambda=1)-H_\HF$, in the present study. Integrating this
expression produces
\begin{equation}
E-E_{0}=\int_{0}^{1}\frac{d\lambda}{\lambda}\langle\Psi(\lambda)|\lambda H_{1}(\lambda
)|\Psi(\lambda)\rangle.
\label{4.4}
\end{equation}
Hence again one is led to a level-shift formula, with $E_{0}$ denoting the
HF ground state energy but now via the coupling constant integration in
Eq.~(\ref{4.4}).

Fetter and Walecka then display the way in which Eq.~(\ref{4.4}), for $N$
electrons, can be reduced to integrations over just two vectors $\mathbf{r}_{1}$ and $\mathbf{r}_{2}$, by means of a time-dependent Green function,
denoted now by $G^{\lambda}(\mathbf{r}_{1} , t;\mathbf{r}_{2}, t^{\prime})$.
This result, shown in their equation (7.31), then
reads
\begin{widetext}
\begin{equation}
E-E_\HF =\pm\frac{i}{2}\int_{0}^{1}\frac{d\lambda}{\lambda}\int d\mathbf{r}_{1}
\lim_{t^{\prime}\to t^{+}}
\lim_{\mathbf{r}_{2}\to\mathbf{r}_{1}}
\left[  \int d\mathbf{y} 
\delta(\mathbf{r}_{1} - \mathbf{y})
\left(i \hbar
\frac{\partial}{\partial t}-h_\HF
(\mathbf{r}_{1},\mathbf{y})\right) 
\Tr G^{\lambda}(\mathbf{y} , t;\mathbf{r}_{2} ,
t^{\prime})\right]  , 
\label{4.5}
\end{equation}
where in place of the kinetic energy operator, in our case the Fock operator
$h_\HF$ appears. As with the level-shift formula in section~\ref{sec:els}, we
now take the differential form of Eq.~(\ref{4.5}) to find the second result
proposed in the present study for the correlation energy density
$\epsilon_{c}(\mathbf{r})$, namely
\begin{equation}
\epsilon_{c}(\mathbf{r}_{1})=\pm\frac{i}{2}\int_{0}^{1}\frac{d\lambda}{\lambda}
\lim_{t^{\prime}\to t^{+}}
\lim_{\mathbf{r}^{\prime}\to\mathbf{r}}
\left[  \int d\mathbf{y} 
\delta(\mathbf{r}_{1} - \mathbf{y})
\left( i\hbar
\frac{\partial}{\partial t}-h_\HF
(\mathbf{r}_{1},\mathbf{y})\right) 
\Tr G^{\lambda}(\mathbf{y} ,
t;\mathbf{r}_{2} , t^{\prime})\right]  . 
\label{4.6a}%
\end{equation}
\end{widetext}
While Eq.~(\ref{4.6a}) is able to furnish a local definition of correlation
energy density, the explicit determination of the Green function $G^{\lambda
}(\mathbf{r}_{1} , t;\mathbf{r}_{2} , t^{\prime})$ represents a
considerable challenge.

It is worth noticing that when the Green function is taken in the HF
approximation, since the operator $\left(i \hbar\frac{\partial}{\partial
t}-h_\HF \right)$ furnishes the inverse of this HF Green function, it follows
that the correlation energy vanishes as it should do, since its definition is
the difference between the exact energy $E$ and the HF value $E_\HF$. This
property has an interesting implication. Let us assume that the correlation
energy density defined in Eq.~(\ref{2.1}) has a dependence on $\mathbf{r}$ which
closely follows the one associated of the HF density (or the exact one )
\cite{Grassi:07}. Then, from definition (\ref{4.6a}) it follows that the HF
correction to the propagator does not contribute at all to the correlation
energy. Therefore, the approximate validity of the correlation energy density
formula proposed in \cite{Grassi:07} means that the contributions to the
correlation energy density coming from the higher order corrections to the Green
function, should approximately follow the behavior of either the total or the HF
density. Therefore, the validity of the proposals of \cite{Grassi:07} for the
correlation energy density, directly implies that the exact (or the HF) density
should have a close relation with the exact or the HF densities. This conclusion
arises because the higher corrections to the HF density turn to be approximately
proportional to the same exact (or HF) densities as implied by the correctness
of the definitions given in \cite{Grassi:07}. This property is supported by the
analytical results of section~\ref{sec:euls} in which a close similarity between
the HF electron density and the correlation energy density emerged. These
general issues are expected to be analyzed in future extensions of the work.

Returning briefly to the theme of the correlation energy density $\epsilon_c
(\br_1 )$, we have written in Eq.~(\ref{4.6a}) a formula involving the Green
function $G^\lambda$ which can be brought into contact with a similar result for
the level shift formula. Let us note that Eq.~(\ref{4.6a}), as well
as Eq.~(\ref{2.1}), can always be rewritten in the form:
\begin{equation}
\epsilon_c (\br_1 ) = \int C(\br_1 ,\br_2 ) d\br_2
\end{equation}
following a similar generalization for kinetic and exchange energy densities,
but now in HF theory, proposed by March and Santamaria
\cite{March:91b,March:06c}. Then the level shift (LS) formula, Eq.~(\ref{2.1})
reads
\begin{equation}
C^{\mathrm{LS}}
(\br_{1} ,\br_2)=\frac{\int\Psi^\ast [H-H_\HF]\Phi_\HF 
d\br_3 \ldots \br_N}{\int\Psi^\ast \Phi_\HF
d\br_1 \ldots d\br_N}.
\label{eq:appast}
\end{equation}
The conventional exact wavefunction (WF) theory reads
\begin{equation}
C^{\mathrm{WF}} (\br_1 , \br_2 ) =
\frac{\int \Psi^\ast H \Psi d\br_3 \ldots d\br_N}{\int\Psi^\ast \Psi
d\br_1 \ldots d\br_N }
-
\frac{\int \Phi_\HF^\ast H \Phi_\HF d\br_3 \ldots d\br_N}{\int\Phi_\HF^\ast \Phi_\HF
d\br_1 \ldots d\br_N } .
\label{eq:appdag}
\end{equation}
While, in general, these forms of $C(\br_1 ,\br_2 )$ are different, they must
all integrate to the same $\int\epsilon_c (\br_1 ) d\br_1$. This is true also for the two
coupling constant integration formulae derivable from Eqs.~(\ref{4.6}) and
(\ref{4.6a}).

What we stress in this Appendix is the $N$-particle character of
Eqs.~(\ref{eq:appast}) and (\ref{eq:appdag}), and the Green function
generalization of $C(\br_1 ,\br_2 )$ following from Eq.~(\ref{4.5}). To date, of
course, analytical progress is restricted to the Moshinsky atom. But since we
have treated this fully in the body of the text, we shall omit further details.

\bibliographystyle{mprsty}
\bibliography{a,b,c,d,e,f,g,h,i,j,k,l,m,n,o,p,q,r,s,t,u,v,w,x,y,z,zzproceedings,Angilella}

\end{document}